\documentclass[aps,physrev,reprint,superscriptaddress]{revtex4-2}

\usepackage{xcolor}
\usepackage{graphicx}
\usepackage{amsmath}

\newcommand{\MnSi}{Mn$_5$Si$_3$}

\begin{document}

\title{Rapid supercurrent decay in Mn$_5$Si$_3$ Josephson junctions}

\author{Arjun Sapkota}
\affiliation{Materials Science, Engineering, and Commercialization Program, Texas State University, San Marcos, Texas 78666, USA}

\author{Kurt Lorenzen}
\affiliation{Department of Physics, Texas State University, San Marcos, Texas 78666, USA}

\author{Tyler Kuhn}
\affiliation{Department of Physics, Texas State University, San Marcos, Texas 78666, USA}

\author{Juan Gomez}
\affiliation{Department of Physics, Texas State University, San Marcos, Texas 78666, USA}

\author{Demet Korucu}
\affiliation{Department of Physics and Astronomy, Michigan State University, East Lansing, Michigan 48824, USA}

\author{Robert M. Klaes}
\affiliation{Department of Physics and Astronomy, Michigan State University, East Lansing, Michigan 48824, USA}

\author{Reza Loloee}
\affiliation{Department of Physics and Astronomy, Michigan State University, East Lansing, Michigan 48824, USA}

\author{Norman O. Birge}
\affiliation{Department of Physics and Astronomy, Michigan State University, East Lansing, Michigan 48824, USA}

\author{Nathan Satchell}
\email{satchell@txstate.edu}
\affiliation{Department of Physics, Texas State University, San Marcos, Texas 78666, USA}
\affiliation{Materials Science, Engineering, and Commercialization Program, Texas State University, San Marcos, Texas 78666, USA}

\vspace{24pt}

\begin{abstract}
Theoretical work predicts that Josephson junctions containing metallic altermagnetic barriers should display $0$--$\pi$ transitions of the critical current as a function of both barrier thickness and temperature, with the decay and oscillation period of the supercurrent depending on the orientation of the crystal axes relative to the transport direction. Motivated by these predictions, and by reports of a compensated magnetic phase attributed to altermagnetism in epitaxial \MnSi{} thin films, we fabricate and measure Nb/Pt/\MnSi{}/Pt/Nb Josephson junctions varying the thickness of the \MnSi{} barrier. The critical current decays as a single exponential over more than four orders of magnitude with decay length $\xi_{\text{Mn}_5\text{Si}_3} = 0.31 \pm 0.03$~nm, shorter than reported for Josephson junctions containing the metallic antiferromagnets FeMn, Cr, and NiMn. The \MnSi{} barrier has an estimated current-perpendicular-to-plane resistivity of $320 \pm 10~\mu\Omega\,$cm. No $0$--$\pi$ transition is resolved at the sampled barrier thicknesses, and the temperature dependence of the critical current of a junction with a 1~nm barrier is smooth and monotonic. We discuss the absence of resolvable transitions in terms of the microstructure of the barrier, its uncertain magnetic phase, and the narrow thickness window imposed by the rapid decay, and identify barriers with well-defined crystalline orientation as the key requirement for future tests of altermagnetic Josephson physics.
\end{abstract}

\maketitle
\newpage

\section{Introduction}
Josephson junctions containing magnetic barriers are a platform for studying the interplay of superconducting and magnetic orders, and are under development for applications in superconducting logic, cryogenic memory, and certain qubit realizations. Of particular interest are transitions of the junction ground state phase between $0$ and $\pi$, firmly established experimentally in ferromagnetic junctions through both the thickness and the temperature dependence of the critical current \cite{Ryazanov_PRL_2001, kontos_2002, 10.1063/5.0195229}. Magnetic barriers with zero net moment, such as antiferromagnets and the recently identified altermagnets~\cite{smejkal_2022}, offer complementary opportunities for the same technologies: they generate no stray fields, they are comparatively insensitive to moderate applied fields, and they avoid the offsets and hysteresis that the remanent magnetization of a ferromagnetic barrier imposes on the magnetic interference pattern.

Theoretically, $0$--$\pi$ transitions are possible in antiferromagnetic and altermagnetic barriers with zero net moment. In antiferromagnets, microscopic calculations predict $0$ or $\pi$ coupling controlled by the atomic scale structure of the barrier, alternating between even and odd numbers of antiferromagnetic layers \cite{Andersen2006}. In altermagnets, metallic barriers are predicted to produce $0$--$\pi$ oscillations of the Josephson coupling as a function of barrier thickness, or of temperature at fixed thickness, with the decay and oscillation period depending on the orientation of the crystal axes relative to the transport direction~\cite{ouassou_2023, zhang_2023, Fukaya2025}. These calculations assume monodomain, crystallographically oriented barriers.

Josephson coupling through metallic antiferromagnets in the current-perpendicular-to-plane (CPP) geometry has been established experimentally in a small number of systems. Bell \textit{et al.} reported the first junctions using $\gamma$-Fe$_{50}$Mn$_{50}$ and found a single exponential decay of the critical current with a characteristic length of order 1~nm \cite{Bell2003}. Comparable behavior has been reported for junctions containing the itinerant antiferromagnet Cr \cite{Weides2009} and the alloy antiferromagnet NiMn \cite{Klaes2023}. Separately, in the lateral junction geometry, Josephson supercurrents have been reported through the altermagnetic candidates RuO$_2$~\cite{prateek2024fabrication} and CrSb~\cite{esin2026josephson}, and through the chiral noncollinear antiferromagnet Mn$_3$Ge ~\cite{jeon2021long}, where the supercurrents are long ranged and have been attributed to spin triplet pairing generated by the chiral magnetic order of the barrier.

\MnSi{} hosts several magnetic phases. Bulk \MnSi{} orders antiferromagnetically below $T_{N2} \approx 100$~K, adopting a noncollinear spin arrangement below $T_{N1} \approx 62$~K in which a large topological Hall response has been reported \cite{surgers_2016}. In epitaxial thin films grown on Si(111), a distinct compensated antiferromagnetic phase has been reported with a N\'eel temperature of $\approx 240$~K, accompanied by a spontaneous anomalous Hall effect consistent with predictions for altermagnetic order \cite{reichlova_2024, kounta_2023}; first principles calculations place this thin film phase among the $d$-wave altermagnetic candidates \cite{smejkal_2022}. Although the magnetic state of our thin \MnSi{} layers deposited without substrate heating is not known, the exploration of Josephson transport in \MnSi{} is a natural first step toward CPP altermagnetic Josephson junctions.

\begin{figure*}[]
\centering
\includegraphics[width=\textwidth]{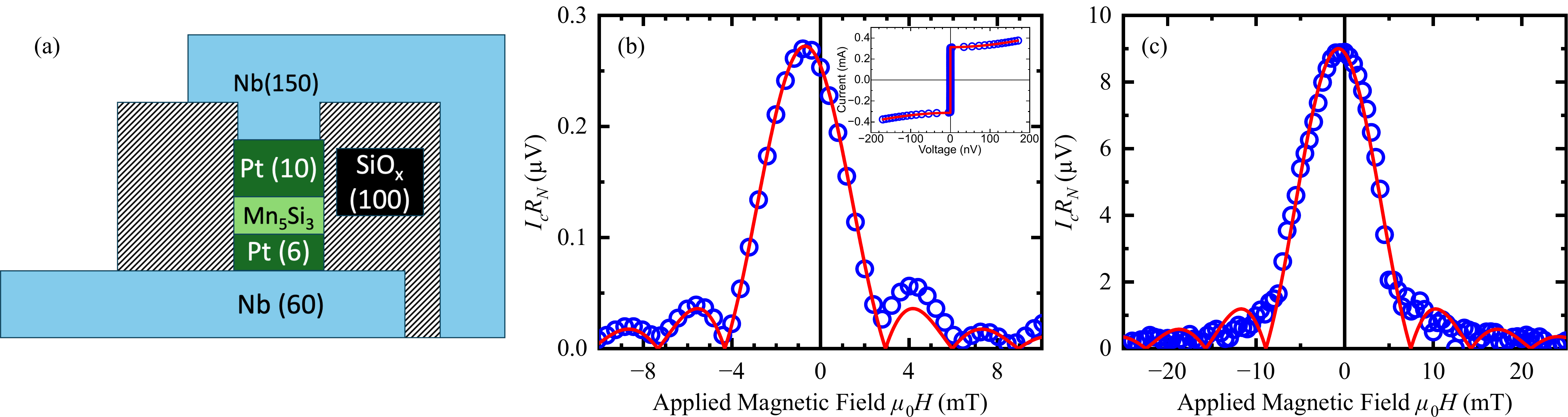}
\caption{(a) Schematic cross section of the Josephson junction device, with layer thicknesses given in (nm) as grown. The thickness of the \MnSi{} layer ranges from 0.5 to 4.5~nm. (b,c) Product of the critical current and normal state resistance, $I_cR_N$, versus applied magnetic field for Josephson junctions with \MnSi{} thicknesses of (b) 2.0~nm and (c) 1.0~nm at 4.2~K. The solid red lines are fits to the data of Eqs.~\ref{eq:Ic} and \ref{eq:phii}. Inset of (b): the $I$--$V$ characteristic of the same junction at zero applied field, together with the fit to Eq.~\ref{eq:VI} from which $I_c$ and $R_N$ are determined.}
\label{fig:1}
\end{figure*}

In this work, we fabricate and measure Nb-based Josephson junctions containing Pt--\MnSi{}--Pt barriers with thickness $0.5~\mathrm{nm} \leq d_{\text{Mn}_5\text{Si}_3} \leq 4.5~\mathrm{nm}$ and test for the predicted $0$--$\pi$ physics. The junctions display Airy-like diffraction patterns centered close to zero applied field, from which we conclude that the barriers carry no significant net magnetization. The critical current decays exponentially with barrier thickness over more than four orders of magnitude with decay length $\xi_{\text{Mn}_5\text{Si}_3} = 0.31 \pm 0.03$~nm, and the temperature dependence of the critical current for the 1~nm \MnSi{} barrier is smooth. Neither measurement displays $0$--$\pi$ signatures, which we attribute to the lack of strong crystalline order of the barrier. We discuss routes toward future experiments employing epitaxial barriers.

\section{Methods}

The films are deposited onto 0.5~mm thick Si substrates with a 1~$\mu$m thick thermal oxide layer using a dc magnetron sputtering system with a base pressure of $2 \times 10^{-8}$~Torr. Growth is performed in approximately 2~mTorr of Ar at typical rates of 0.5~nm\,s$^{-1}$ for Nb, 0.15~nm\,s$^{-1}$ for Pt, and 0.06~nm\,s$^{-1}$ for \MnSi{}, with the substrates held at approximately $-20\,^{\circ}$C during growth. The purity of the Nb target is 99.99\%, and the Pt and \MnSi{} (stoichiometric compound) targets are 99.95\%. Deposition rates are calibrated using an \textit{in situ} quartz crystal thickness monitor and confirmed using x-ray reflectivity measurements on reference films (see Supplemental Material Fig. S1~\cite{SI}). Growth of the bottom 60~nm Nb electrode, 6~nm Pt buffer layer, \MnSi{} barrier layer of varying thickness, and 10~nm Pt capping layer is performed without breaking vacuum. The films are patterned into circular Josephson junctions of diameter 3 and 6~$\mu$m using standard photolithography and ion milling methods, described in previous work~\cite{Wang_2012}. In the final stage of fabrication, the samples are loaded back into the sputtering system and approximately 3~nm of the Pt capping layer is ion milled \textit{in situ}, recovering a clean interface before deposition of the top 150~nm Nb electrode. A schematic of the device structure is shown in Figure~\ref{fig:1} (a).

Electrical transport is performed in two systems: a 4.2~K low-noise potentiometer comparator setup described in Ref.~\cite{Glick2017triplet}, and a variable temperature Quantum Design DynaCool PPMS equipped with Lakeshore M81 measurement electronics. The thickness dependence of the critical current is measured at 4.2~K, with the two systems together covering a large range of critical current: junctions with the largest critical currents ($d \leq 2$~nm) are measured in the PPMS, while the smaller critical currents of the thickest barriers ($d \geq 2$~nm) require the lower noise comparator circuit. The $d = 2$~nm sample is measured in both systems, and we report the lower noise measurements. Temperature dependent measurements are performed in the PPMS. The magnetic field during transport measurements is applied parallel to the sample plane.

\section{Results}

We measure current-voltage ($I$--$V$) characteristics of the junctions as a function of applied magnetic field at a fixed temperature. The $I$--$V$ curves indicate our junctions are overdamped, and can therefore be described by~\cite{barone1982physics}
\begin{equation}
V(I) =
\begin{cases}
\dfrac{I}{|I|}\, R_N \sqrt{I^2 - I_c^2} & \text{for } |I| \geq I_c,\\[1ex]
0 & \text{for } |I| < I_c,
\end{cases}
\label{eq:VI}
\end{equation}
where $I_c$ is the critical current through the junction and $R_N$ is the normal state resistance across the junction. Fitting Eq.~\ref{eq:VI} to each measured characteristic yields $I_c$ and $R_N$; a representative characteristic and fit are shown in the inset of Figure~\ref{fig:1}(b). The product $I_cR_N$ is a useful area-independent quantity that we report as the primary junction metric in this work. We also report the area resistance product, $AR_N$, calculated from the nominal junction area.

When a magnetic field is applied parallel to the sample plane, perpendicular to the current flow across the junction, the critical current shows a ``Fraunhofer'' interference pattern, as shown in Figure~\ref{fig:1}(b,c). Since the junctions are circular, the dependence of $I_cR_N$ on applied magnetic field can be described by the Airy function~\cite{barone1982physics}:
\begin{equation}
I_cR_N ({\Phi}) = I_cR_N(\text{max}) \left| \frac{2J_1 \left(\pi \frac{\Phi}{\Phi_0} \right)}{\pi \frac{\Phi}{\Phi_0}} \right|,
\label{eq:Ic}
\end{equation}
where $I_cR_N(\text{max})$ is the maximum critical current product through the junction, $J_1$ is a Bessel function of the first kind, $\Phi_0$ is the flux quantum ($h/2e$) and $\Phi$ is the flux through the junction, given by~\cite{barone1982physics}
\begin{equation}
\begin{aligned}
\Phi &= \mu_0 (H_{\text{app}} - H_{\text{shift}}) w \Bigg[
\lambda_L^{\text{bottom}} \tanh \left( \frac{d_S^{\text{bottom}}}{2\lambda_L^{\text{bottom}}} \right) \\
&\quad + \lambda_L^{\text{top}} \tanh \left( \frac{d_S^{\text{top}}}{2\lambda_L^{\text{top}}} \right) + d
\Bigg],
\label{eq:phii}
\end{aligned}
\end{equation}
where $w$ is the width of the junction, $H_{\text{app}}$ is the applied field, $H_{\text{shift}}$ is the amount of field by which $I_cR_N(\text{max})$ is shifted from $H=0$, $d_S$ is the thickness of the superconducting electrodes, and $d$ is the total barrier thickness (Mn$_5$Si$_3$ and Pt layers). Since both electrodes are Nb, we set the London penetration depth, $\lambda_L^{\text{bottom}} = \lambda_L^{\text{top}} = 100$~nm~\cite{quarterman2020distortions}.

Fits of Eqs.~\ref{eq:Ic} and \ref{eq:phii} describe the measured interference patterns well, with the amplitude $I_cR_N(\text{max})$, the width $w$, and the offset $H_{\text{shift}}$ as free parameters. The fits to Figure~\ref{fig:1}(b,c) give best fit $w$ of 7 and 3~$\mu$m respectively, the latter matching the nominal photolithographic diameter of 3~$\mu$m, while the former exceeds it. The origin of this discrepancy is unknown and left for future investigation. Because $I_cR_N$ is independent of junction area, it does not affect the decay length or the temperature dependence, which are our main results. Averaged over all junctions measured at 4.2~K, the offset field is $|\mu_0 H_{\text{shift}}| \approx 0.7$~mT. This small, thickness independent offset most likely originates from trapped flux in the superconducting solenoid used to apply the magnetic field, rather than from any in-plane magnetization of the barrier, consistent with our previous work~\cite{sapkota_2026}. The centered interference patterns are therefore consistent with a barrier carrying no significant net magnetization, in contrast to the shifted and hysteretic patterns characteristic of ferromagnetic Josephson junctions~\cite{10.1063/5.0195229}.

\begin{figure}[]
    \centering
    \includegraphics[width=\linewidth]{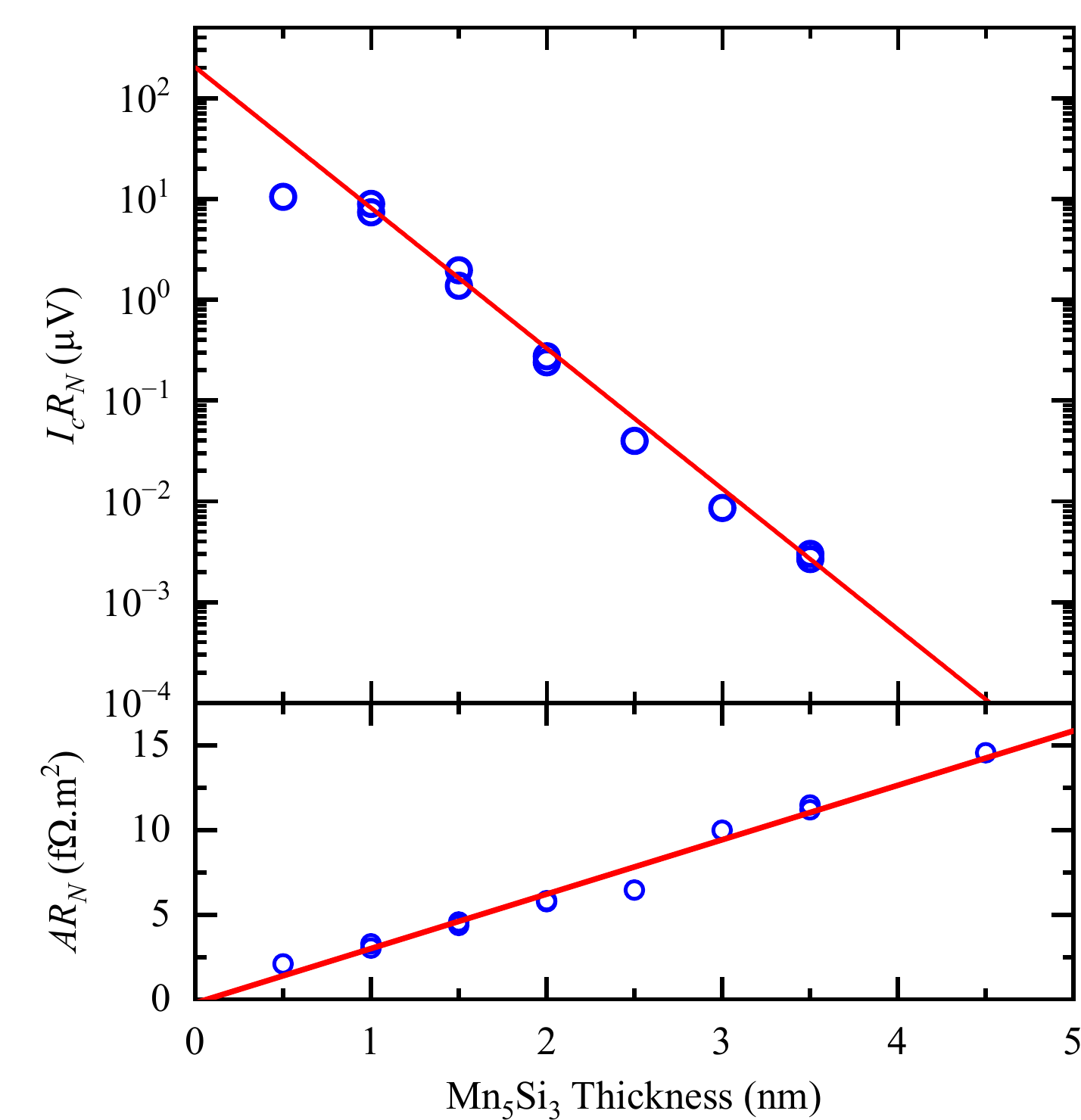}
    \caption{Thickness dependence of supercurrent in Mn$_5$Si$_3$ Josephson junctions. \textbf{Top:} Product of the critical current and normal state resistance, $I_cR_N$ versus Mn$_5$Si$_3$ thickness at 4.2~K. The line is a best fit to Eq.~\ref{eq:decay} over the range 1--3.5~nm with decay length $\xi_{\text{Mn}_5\text{Si}_3} = 0.31 \pm 0.03$~nm. \textbf{Bottom:} Area resistance product, $AR_N$, for the junctions, where the best fit to Eq.~\ref{eq:4} indicates the Mn$_5$Si$_3$ resistivity, $\rho=320 \pm 10~\mu\Omega\,\mathrm{cm}$. Each data point represents one Josephson junction and the uncertainty in determining $I_cR_N$ and $AR_N$ is smaller than the data points.}
    \label{fig:2}
 \end{figure}

\begin{figure}[h]
    \centering
    \includegraphics[width=\linewidth]{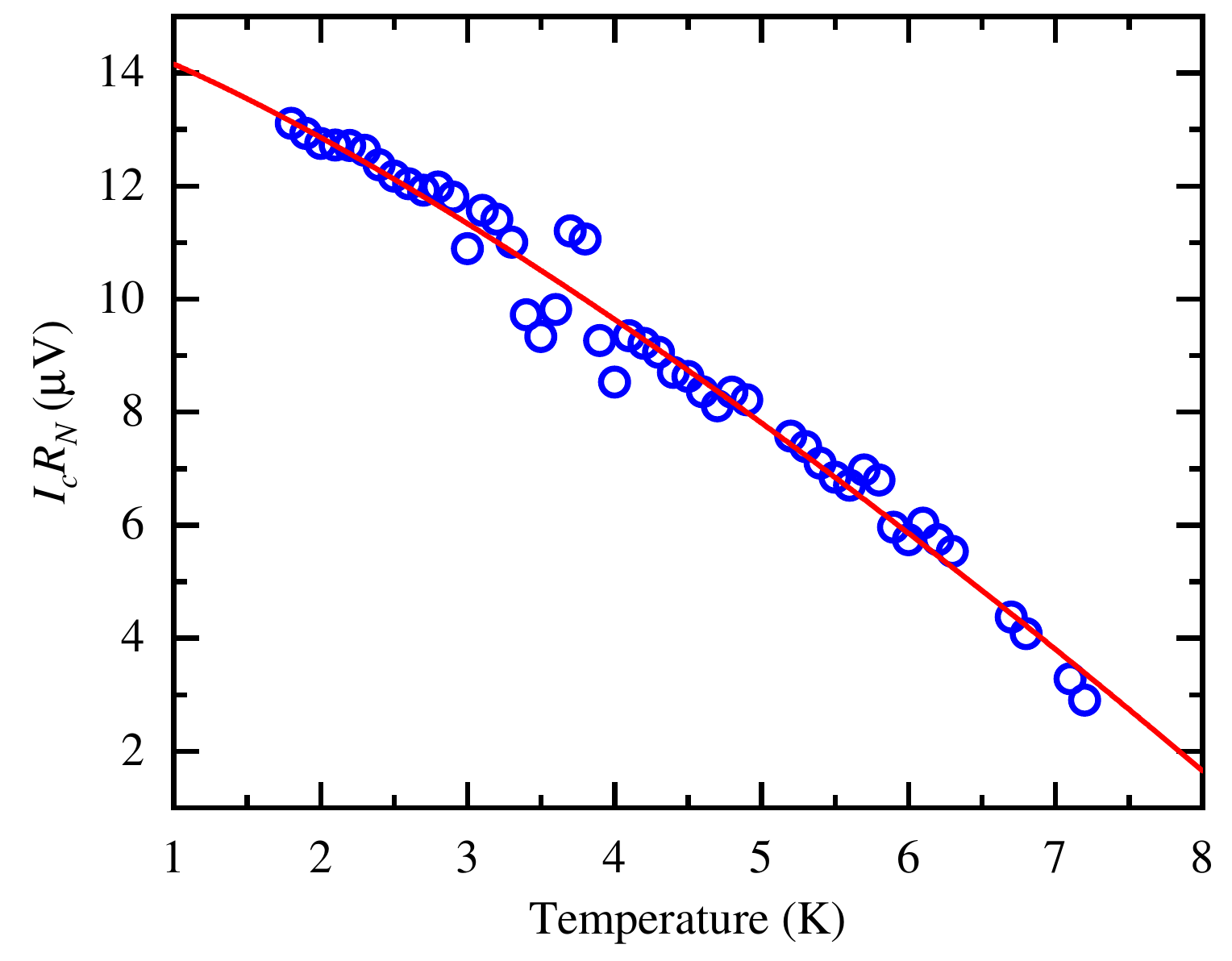}
    \caption{Product of the critical current and normal state resistance, $I_cR_N$ versus temperature for a Josephson junction with Mn$_5$Si$_3$ thickness of 1.0~nm. The critical temperature is approximately $8.8$~K, and the trend of $I_cR_N(T)$ is well described by the phenomenological expression, Eq.~\ref{eq:J_c(T)}. Each data point represents $I_cR_N$ extracted at one temperature and the uncertainty in determining $I_cR_N$ is smaller than the data points.}
    \label{fig:3}
\end{figure}

For each junction, we extract $I_cR_N(\text{max})$ measured at 4.2~K. The thickness dependence is shown in Figure~\ref{fig:2} (top). Between 1~nm and 3.5~nm, the data follow a single exponential decay spanning more than four orders of magnitude in $I_cR_N$, while the 0.5~nm junction falls below the trend. A likely explanation is the crossover into the large junction limit. We estimate the Josephson penetration depth from $\lambda_J=\sqrt{\hbar/[2e\mu_0(2\lambda_L+d)J_c]}$~\cite{barone1982physics}, where $J_c = I_c/A$ is evaluated from the nominal junction area; this gives $\lambda_J\approx0.5~\mu$m for the 0.5~nm junction, comparable to the junction dimension. In this regime, screening by the self field of the Josephson current suppresses the measured critical current below its true value~\cite{barone1982physics}. The 1~nm junction, by contrast, lies on the exponential trend, and we restrict the decay fitting to the range 1--3.5~nm. A junction with $d_{\text{Mn}_5\text{Si}_3} = 4.5$~nm was also measured, but displayed no resolvable supercurrent at the 10~nA current step of the low noise measurement. The $I$--$V$ characteristic of this junction is linear, and provides the $AR_N$ value included in Figure~\ref{fig:2} (bottom).

We describe the decay of $I_cR_N$ over the thickness range 1--3.5~nm using the exponential:
\begin{equation}
  I_cR_N = I_cR_N(0) \exp\left(\frac{-d_{\text{Mn}_5\text{Si}_3}}{\xi_{\text{Mn}_5\text{Si}_3}}\right),
\label{eq:decay}
\end{equation}
where $\xi_{\text{Mn}_5\text{Si}_3}$ is the effective decay length inside the Mn$_5$Si$_3$ and $d_{\text{Mn}_5\text{Si}_3}$ is the thickness. $I_cR_N(0)$ is an extrapolation to zero barrier thickness and is not a physical critical current~\cite{10.1063/5.0195229}. The best fit presented gives $\xi_{\text{Mn}_5\text{Si}_3} = 0.31 \pm 0.03$~nm and $I_cR_N(0) = 200 \pm 70~\mu$V.

Figure~\ref{fig:2} (bottom) shows the area resistance product versus the Mn$_5$Si$_3$ thickness. The resistivity, $\rho_{\mathrm{Mn_5Si_3}}$, in the CPP direction and the combined resistance of the interfaces and Pt layers, $2AR$, can be estimated as,
\begin{equation}
A R_N = \rho_{\mathrm{Mn_5Si_3}}\, d_{\mathrm{Mn_5Si_3}} + 2AR.
\label{eq:4}
\end{equation}
The best fit to our data gives $\rho_{\mathrm{Mn_5Si_3}} = 320 \pm 10~\mu\Omega\,\mathrm{cm}$ and an intercept consistent with zero, $2AR = -0.2 \pm 0.5$~f$\Omega$m$^2$, which bounds the combined contribution of the interfaces and Pt layers to $2AR \lesssim 0.5$~f$\Omega$m$^2$.

Next we measure the temperature dependence of $I_cR_N$ in the 1~nm sample by collecting the interference pattern at each temperature and extracting $I_cR_N(T)$, Figure~\ref{fig:3}. $I_cR_N$ is largest at the lowest temperature and decreases smoothly and monotonically with increasing temperature; we observe no dip or other nonmonotonic feature of the kind predicted at a temperature driven $0$--$\pi$ transition~\cite{ouassou_2023}. In the absence of a $0$--$\pi$ transition, $I_cR_N(T)$ can be described phenomenologically as
\begin{equation}
I_cR_N(T) = I_cR_N(0~\text{K}) \left[ 1 - \left( \frac{T}{T_{c}} \right)^n \right],
\label{eq:J_c(T)}
\end{equation}
where $I_cR_N(0~\text{K})$ is the extrapolated critical current product at zero temperature, $T_{c}$ is the critical temperature, and $n$ is an empirically determined exponent. The best fit gives $I_cR_N(0~\text{K}) = 15.0 \pm 0.7~\mu$V, $T_{c} = 8.8 \pm 0.2$~K, and $n = 1.30 \pm 0.14$.

\section{Discussion}

The calculations of Ref.~\cite{ouassou_2023} make three concrete predictions for Josephson junctions with metallic altermagnetic barriers: $0$--$\pi$ oscillations of the coupling as a function of barrier length, possible $0$--$\pi$ transitions as a function of temperature, and a strong dependence of both the decay and the oscillation period on the orientation of the crystal axes relative to the transport direction. Our junctions show none of the oscillatory signatures. We consider the most likely explanations in turn.

We first consider the microstructure of the barrier. In ferromagnetic Josephson junctions, the exchange splitting responsible for the $0$--$\pi$ oscillations persists irrespective of the crystalline order or orientation of the barrier. The altermagnetic spin splitting is qualitatively different: it alternates in sign with the orientation of the crystal axes \cite{smejkal_2022,ouassou_2023}. The predicted $0$--$\pi$ physics may therefore require a barrier with a well defined crystalline orientation, and the calculations of Ref.~\cite{ouassou_2023} assume a monodomain, crystallographically oriented barrier. In our junctions, the current flows perpendicular to the film plane through a barrier without a strong crystalline microstructure. An analogous argument holds for the parity dependent $0$ and $\pi$ coupling predicted for atomically ordered antiferromagnetic barriers \cite{Andersen2006}, which is unobserved in junctions containing polycrystalline Cr \cite{Weides2009}.

A second consideration is that the magnetic order of our \MnSi{} barriers may not coincide with the compensated phase reported in epitaxial \MnSi{} on Si(111) and attributed to altermagnetism \cite{reichlova_2024}. Our interference patterns indicate that the net in-plane magnetization of the \MnSi{} barriers is small, but provide no information on their magnetic ordering. Since the magnetic state of \MnSi{} is known to be complex, our barrier layers may instead host the bulk-like noncollinear antiferromagnetic phase or another phase that does not realize the collinear spin splitting from which the predicted $0$--$\pi$ oscillations arise \cite{ouassou_2023}.

A third consideration is the range of barrier thicknesses over which the supercurrent can be measured, compared to the length scale of the predicted oscillations. The oscillation period for \MnSi{} is not known and the very short decay length compresses the accessible window: $I_cR_N$ falls by more than four orders of magnitude between 1 and 3.5~nm, and no supercurrent is resolvable at 4.5~nm. If the oscillation period is long compared to this window, the junctions run out of measurable supercurrent before the first $0$--$\pi$ crossing is reached; a period shorter than our 0.5~nm thickness increment may likewise escape detection. The single-exponential behavior of Figure~\ref{fig:2} therefore excludes only a resolvable $0$--$\pi$ crossing at the sampled thicknesses; it cannot exclude a crossing between those thickness or beyond the accessible window. A similar limitation applies to the temperature dependence: the predicted nonmonotonic feature in $I_cR_N(T)$ occurs only for barrier thicknesses close to a $0$--$\pi$ boundary \cite{ouassou_2023}. The smooth $I_cR_N(T)$ of the 1~nm junction indicates that this thickness does not lie near such a boundary, but we cannot exclude temperature driven transitions more generally.

\begin{table}
\caption{Characteristic supercurrent decay lengths $\xi$ reported for Josephson junctions containing metallic antiferromagnetic barriers, defined through $J_c \propto \exp(-d/\xi)$, together with the value for \MnSi{} measured in this work.}
\begin{ruledtabular}
\begin{tabular}{lll}
Barrier & $\xi$ (nm) & Ref. \\
\hline
$\gamma$-Fe$_{50}$Mn$_{50}$ & 1.2\footnotemark[1] & \cite{Bell2003} \\
Cr & 2.1--3.4 & \cite{Weides2009} \\
Ni$_{41}$Mn$_{59}$ & $1.03 \pm 0.05$ & \cite{Klaes2023} \\
\MnSi{} & $0.31 \pm 0.03$ & This work \\
\end{tabular}
\end{ruledtabular}
\footnotetext[1]{Reference~\cite{Bell2003} quotes $\xi = 2.4$~nm using the convention $J_c \propto \exp(-2d/\xi)$; the value is converted here to the single exponential convention used throughout this paper.}
\label{tab:af}
\end{table}

Finally, we compare the measured decay length to junctions containing metallic antiferromagnets, collected in Table~\ref{tab:af}. The value $\xi_{\text{Mn}_5\text{Si}_3} = 0.31 \pm 0.03$~nm is a factor of 3 to 11 shorter than reported for NiMn, $\gamma$-Fe$_{50}$Mn$_{50}$, and Cr \cite{Klaes2023,Bell2003,Weides2009}.  

None of these considerations diminishes the central experimental point: in junctions with a well defined CPP current direction and sub-nanometer control of the barrier thickness, Josephson transport through \MnSi{} without strong crystalline order shows a rapid, single exponential supercurrent decay with no sign of the predicted oscillations. Moreover, none of the considerations is specific to \MnSi{}: each applies to any candidate altermagnet barrier, and together they set a single requirement for future experiments, a barrier with well defined crystalline orientation. Orientation fixes the crystal axes relative to the transport direction~\cite{ouassou_2023}. For \MnSi{}, epitaxial growth is also the route by which the compensated phase is stabilized: films on Si(111) show the relevant order \cite{reichlova_2024, kounta_2023}, but have not yet been incorporated in a heterostructure compatible with Josephson junction fabrication, and considerable materials development will be required to realize the necessary superconductor/altermagnet stacks. The junction process and benchmark transport values reported here provide the starting point for that effort.

\section{Conclusions}

We have fabricated Nb/Pt/\MnSi{}/Pt/Nb Josephson junctions containing \MnSi{} barriers with thicknesses between 0.5 and 4.5~nm and without strong crystalline order. The junctions display ``Fraunhofer'' interference patterns centered close to zero applied field, indicating a barrier carrying no significant net magnetization. The critical current decays as a single exponential over more than four orders of magnitude with decay length $\xi_{\text{Mn}_5\text{Si}_3} = 0.31 \pm 0.03$~nm, a factor of three or more shorter than reported for Josephson junctions containing the metallic antiferromagnets FeMn, Cr, and NiMn, and no supercurrent is resolvable at 4.5~nm. From the area resistance product we obtain the resistivity of the barrier in the transport geometry of the devices, $320 \pm 10~\mu\Omega\,$cm, with negligible interface resistance. No $0$--$\pi$ transition is resolved at the measured barrier thicknesses, while the smooth and monotonic temperature dependence of the critical current of the 1~nm junction indicates that this thickness does not lie near a temperature-driven transition. The predicted transitions now await tests in junctions containing crystallographically oriented barriers.

\section*{Data Availability Statement}

The data that support the findings of this article are publicly available in the Texas Data Repository~\cite{Data}.

\begin{acknowledgments}
We acknowledge experimental assistance through the Analysis Research Service Center from Sam Cantrell and Casey Smith, and fabrication assistance using the Keck Microfabrication Facility from Baokang Bi. We acknowledge support from new faculty startup funding made available by Texas State University.
\end{acknowledgments}

\bibliography{Refs}

\begin{thebibliography}{24}%
\makeatletter
\providecommand \@ifxundefined [1]{%
 \@ifx{#1\undefined}
}%
\providecommand \@ifnum [1]{%
 \ifnum #1\expandafter \@firstoftwo
 \else \expandafter \@secondoftwo
 \fi
}%
\providecommand \@ifx [1]{%
 \ifx #1\expandafter \@firstoftwo
 \else \expandafter \@secondoftwo
 \fi
}%
\providecommand \natexlab [1]{#1}%
\providecommand \enquote  [1]{``#1''}%
\providecommand \bibnamefont  [1]{#1}%
\providecommand \bibfnamefont [1]{#1}%
\providecommand \citenamefont [1]{#1}%
\providecommand \href@noop [0]{\@secondoftwo}%
\providecommand \href [0]{\begingroup \@sanitize@url \@href}%
\providecommand \@href[1]{\@@startlink{#1}\@@href}%
\providecommand \@@href[1]{\endgroup#1\@@endlink}%
\providecommand \@sanitize@url [0]{\catcode `\\12\catcode `\$12\catcode `\&12\catcode `\#12\catcode `\^12\catcode `\_12\catcode `\%12\relax}%
\providecommand \@@startlink[1]{}%
\providecommand \@@endlink[0]{}%
\providecommand \url  [0]{\begingroup\@sanitize@url \@url }%
\providecommand \@url [1]{\endgroup\@href {#1}{\urlprefix }}%
\providecommand \urlprefix  [0]{URL }%
\providecommand \Eprint [0]{\href }%
\providecommand \doibase [0]{https://doi.org/}%
\providecommand \selectlanguage [0]{\@gobble}%
\providecommand \bibinfo  [0]{\@secondoftwo}%
\providecommand \bibfield  [0]{\@secondoftwo}%
\providecommand \translation [1]{[#1]}%
\providecommand \BibitemOpen [0]{}%
\providecommand \bibitemStop [0]{}%
\providecommand \bibitemNoStop [0]{.\EOS\space}%
\providecommand \EOS [0]{\spacefactor3000\relax}%
\providecommand \BibitemShut  [1]{\csname bibitem#1\endcsname}%
\let\auto@bib@innerbib\@empty
\bibitem [{\citenamefont {Ryazanov}\ \emph {et~al.}(2001)\citenamefont {Ryazanov}, \citenamefont {Oboznov}, \citenamefont {Rusanov}, \citenamefont {Veretennikov}, \citenamefont {Golubov},\ and\ \citenamefont {Aarts}}]{Ryazanov_PRL_2001}%
  \BibitemOpen
  \bibfield  {author} {\bibinfo {author} {\bibfnamefont {V.~V.}\ \bibnamefont {Ryazanov}}, \bibinfo {author} {\bibfnamefont {V.~A.}\ \bibnamefont {Oboznov}}, \bibinfo {author} {\bibfnamefont {A.~Y.}\ \bibnamefont {Rusanov}}, \bibinfo {author} {\bibfnamefont {A.~V.}\ \bibnamefont {Veretennikov}}, \bibinfo {author} {\bibfnamefont {A.~A.}\ \bibnamefont {Golubov}},\ and\ \bibinfo {author} {\bibfnamefont {J.}~\bibnamefont {Aarts}},\ }\bibfield  {title} {\bibinfo {title} {Coupling of {T}wo {S}uperconductors through a {F}erromagnet: {E}vidence for a $\ensuremath{\pi}$ {J}unction},\ }\href {https://doi.org/10.1103/PhysRevLett.86.2427} {\bibfield  {journal} {\bibinfo  {journal} {Phys. Rev. Lett.}\ }\textbf {\bibinfo {volume} {86}},\ \bibinfo {pages} {2427} (\bibinfo {year} {2001})}\BibitemShut {NoStop}%
\bibitem [{\citenamefont {Kontos}\ \emph {et~al.}(2002)\citenamefont {Kontos}, \citenamefont {Aprili}, \citenamefont {Lesueur}, \citenamefont {Genêt}, \citenamefont {Stephanidis},\ and\ \citenamefont {Boursier}}]{kontos_2002}%
  \BibitemOpen
  \bibfield  {author} {\bibinfo {author} {\bibfnamefont {T.}~\bibnamefont {Kontos}}, \bibinfo {author} {\bibfnamefont {M.}~\bibnamefont {Aprili}}, \bibinfo {author} {\bibfnamefont {J.}~\bibnamefont {Lesueur}}, \bibinfo {author} {\bibfnamefont {F.}~\bibnamefont {Genêt}}, \bibinfo {author} {\bibfnamefont {B.}~\bibnamefont {Stephanidis}},\ and\ \bibinfo {author} {\bibfnamefont {R.}~\bibnamefont {Boursier}},\ }\bibfield  {title} {\bibinfo {title} {{J}osephson junction through a thin ferromagnetic layer: negative coupling.},\ }\href {https://doi.org/10.1103/{PhysRevLett}.89.137007} {\bibfield  {journal} {\bibinfo  {journal} {Phys. Rev. Lett.}\ }\textbf {\bibinfo {volume} {89}},\ \bibinfo {pages} {137007} (\bibinfo {year} {2002})}\BibitemShut {NoStop}%
\bibitem [{\citenamefont {Birge}\ and\ \citenamefont {Satchell}(2024)}]{10.1063/5.0195229}%
  \BibitemOpen
  \bibfield  {author} {\bibinfo {author} {\bibfnamefont {N.~O.}\ \bibnamefont {Birge}}\ and\ \bibinfo {author} {\bibfnamefont {N.}~\bibnamefont {Satchell}},\ }\bibfield  {title} {\bibinfo {title} {{Ferromagnetic materials for {J}osephson $\pi$ junctions}},\ }\href {https://doi.org/10.1063/5.0195229} {\bibfield  {journal} {\bibinfo  {journal} {APL Mater.}\ }\textbf {\bibinfo {volume} {12}},\ \bibinfo {pages} {041105} (\bibinfo {year} {2024})}\BibitemShut {NoStop}%
\bibitem [{\citenamefont {\ifmmode~\check{S}\else \v{S}\fi{}mejkal}\ \emph {et~al.}(2022)\citenamefont {\ifmmode~\check{S}\else \v{S}\fi{}mejkal}, \citenamefont {Sinova},\ and\ \citenamefont {Jungwirth}}]{smejkal_2022}%
  \BibitemOpen
  \bibfield  {author} {\bibinfo {author} {\bibfnamefont {L.}~\bibnamefont {\ifmmode~\check{S}\else \v{S}\fi{}mejkal}}, \bibinfo {author} {\bibfnamefont {J.}~\bibnamefont {Sinova}},\ and\ \bibinfo {author} {\bibfnamefont {T.}~\bibnamefont {Jungwirth}},\ }\bibfield  {title} {\bibinfo {title} {Emerging {R}esearch {L}andscape of {A}ltermagnetism},\ }\href {https://doi.org/10.1103/PhysRevX.12.040501} {\bibfield  {journal} {\bibinfo  {journal} {Phys. Rev. X}\ }\textbf {\bibinfo {volume} {12}},\ \bibinfo {pages} {040501} (\bibinfo {year} {2022})}\BibitemShut {NoStop}%
\bibitem [{\citenamefont {Andersen}\ \emph {et~al.}(2006)\citenamefont {Andersen}, \citenamefont {Bobkova}, \citenamefont {Hirschfeld},\ and\ \citenamefont {Barash}}]{Andersen2006}%
  \BibitemOpen
  \bibfield  {author} {\bibinfo {author} {\bibfnamefont {B.~M.}\ \bibnamefont {Andersen}}, \bibinfo {author} {\bibfnamefont {I.~V.}\ \bibnamefont {Bobkova}}, \bibinfo {author} {\bibfnamefont {P.~J.}\ \bibnamefont {Hirschfeld}},\ and\ \bibinfo {author} {\bibfnamefont {Y.~S.}\ \bibnamefont {Barash}},\ }\bibfield  {title} {\bibinfo {title} {$0\ensuremath{-}\ensuremath{\pi}$ {Transitions in Josephson Junctions with Antiferromagnetic I}nterlayers},\ }\href {https://doi.org/10.1103/PhysRevLett.96.117005} {\bibfield  {journal} {\bibinfo  {journal} {Phys. Rev. Lett.}\ }\textbf {\bibinfo {volume} {96}},\ \bibinfo {pages} {117005} (\bibinfo {year} {2006})}\BibitemShut {NoStop}%
\bibitem [{\citenamefont {Ouassou}\ \emph {et~al.}(2023)\citenamefont {Ouassou}, \citenamefont {Brataas},\ and\ \citenamefont {Linder}}]{ouassou_2023}%
  \BibitemOpen
  \bibfield  {author} {\bibinfo {author} {\bibfnamefont {J.~A.}\ \bibnamefont {Ouassou}}, \bibinfo {author} {\bibfnamefont {A.}~\bibnamefont {Brataas}},\ and\ \bibinfo {author} {\bibfnamefont {J.}~\bibnamefont {Linder}},\ }\bibfield  {title} {\bibinfo {title} {dc {J}osephson {E}ffect in {A}ltermagnets},\ }\href {https://doi.org/10.1103/PhysRevLett.131.076003} {\bibfield  {journal} {\bibinfo  {journal} {Phys. Rev. Lett.}\ }\textbf {\bibinfo {volume} {131}},\ \bibinfo {pages} {076003} (\bibinfo {year} {2023})}\BibitemShut {NoStop}%
\bibitem [{\citenamefont {Zhang}\ \emph {et~al.}(2024)\citenamefont {Zhang}, \citenamefont {Hu},\ and\ \citenamefont {Neupert}}]{zhang_2023}%
  \BibitemOpen
  \bibfield  {author} {\bibinfo {author} {\bibfnamefont {S.-B.}\ \bibnamefont {Zhang}}, \bibinfo {author} {\bibfnamefont {L.-H.}\ \bibnamefont {Hu}},\ and\ \bibinfo {author} {\bibfnamefont {T.}~\bibnamefont {Neupert}},\ }\bibfield  {title} {\bibinfo {title} {Finite-momentum {C}ooper pairing in proximitized altermagnets},\ }\href {https://doi.org/10.1038/s41467-024-45951-3} {\bibfield  {journal} {\bibinfo  {journal} {Nat. Commun.}\ }\textbf {\bibinfo {volume} {15}},\ \bibinfo {pages} {1801} (\bibinfo {year} {2024})}\BibitemShut {NoStop}%
\bibitem [{\citenamefont {Fukaya}\ \emph {et~al.}(2025)\citenamefont {Fukaya}, \citenamefont {Maeda}, \citenamefont {Yada}, \citenamefont {Cayao}, \citenamefont {Tanaka},\ and\ \citenamefont {Lu}}]{Fukaya2025}%
  \BibitemOpen
  \bibfield  {author} {\bibinfo {author} {\bibfnamefont {Y.}~\bibnamefont {Fukaya}}, \bibinfo {author} {\bibfnamefont {K.}~\bibnamefont {Maeda}}, \bibinfo {author} {\bibfnamefont {K.}~\bibnamefont {Yada}}, \bibinfo {author} {\bibfnamefont {J.}~\bibnamefont {Cayao}}, \bibinfo {author} {\bibfnamefont {Y.}~\bibnamefont {Tanaka}},\ and\ \bibinfo {author} {\bibfnamefont {B.}~\bibnamefont {Lu}},\ }\bibfield  {title} {\bibinfo {title} {Josephson effect and odd-frequency pairing in superconducting junctions with unconventional magnets},\ }\href {https://doi.org/10.1103/PhysRevB.111.064502} {\bibfield  {journal} {\bibinfo  {journal} {Phys. Rev. B}\ }\textbf {\bibinfo {volume} {111}},\ \bibinfo {pages} {064502} (\bibinfo {year} {2025})}\BibitemShut {NoStop}%
\bibitem [{\citenamefont {Bell}\ \emph {et~al.}(2003)\citenamefont {Bell}, \citenamefont {Tarte}, \citenamefont {Burnell}, \citenamefont {Leung}, \citenamefont {Kang},\ and\ \citenamefont {Blamire}}]{Bell2003}%
  \BibitemOpen
  \bibfield  {author} {\bibinfo {author} {\bibfnamefont {C.}~\bibnamefont {Bell}}, \bibinfo {author} {\bibfnamefont {E.~J.}\ \bibnamefont {Tarte}}, \bibinfo {author} {\bibfnamefont {G.}~\bibnamefont {Burnell}}, \bibinfo {author} {\bibfnamefont {C.~W.}\ \bibnamefont {Leung}}, \bibinfo {author} {\bibfnamefont {D.-J.}\ \bibnamefont {Kang}},\ and\ \bibinfo {author} {\bibfnamefont {M.~G.}\ \bibnamefont {Blamire}},\ }\bibfield  {title} {\bibinfo {title} {Proximity and {J}osephson effects in superconductor/antiferromagnetic {N}b$/\gamma-${F}e$_{50}${Mn}$_{50}$ heterostructures},\ }\href {https://doi.org/10.1103/PhysRevB.68.144517} {\bibfield  {journal} {\bibinfo  {journal} {Phys. Rev. B}\ }\textbf {\bibinfo {volume} {68}},\ \bibinfo {pages} {144517} (\bibinfo {year} {2003})}\BibitemShut {NoStop}%
\bibitem [{\citenamefont {Weides}\ \emph {et~al.}(2009)\citenamefont {Weides}, \citenamefont {Disch}, \citenamefont {Kohlstedt},\ and\ \citenamefont {B\"urgler}}]{Weides2009}%
  \BibitemOpen
  \bibfield  {author} {\bibinfo {author} {\bibfnamefont {M.}~\bibnamefont {Weides}}, \bibinfo {author} {\bibfnamefont {M.}~\bibnamefont {Disch}}, \bibinfo {author} {\bibfnamefont {H.}~\bibnamefont {Kohlstedt}},\ and\ \bibinfo {author} {\bibfnamefont {D.~E.}\ \bibnamefont {B\"urgler}},\ }\bibfield  {title} {\bibinfo {title} {Observation of {J}osephson coupling through an interlayer of antiferromagnetically ordered chromium},\ }\href {https://doi.org/10.1103/PhysRevB.80.064508} {\bibfield  {journal} {\bibinfo  {journal} {Phys. Rev. B}\ }\textbf {\bibinfo {volume} {80}},\ \bibinfo {pages} {064508} (\bibinfo {year} {2009})}\BibitemShut {NoStop}%
\bibitem [{\citenamefont {Klaes}\ \emph {et~al.}(2023)\citenamefont {Klaes}, \citenamefont {Loloee},\ and\ \citenamefont {Birge}}]{Klaes2023}%
  \BibitemOpen
  \bibfield  {author} {\bibinfo {author} {\bibfnamefont {R.~M.}\ \bibnamefont {Klaes}}, \bibinfo {author} {\bibfnamefont {R.}~\bibnamefont {Loloee}},\ and\ \bibinfo {author} {\bibfnamefont {N.~O.}\ \bibnamefont {Birge}},\ }\bibfield  {title} {\bibinfo {title} {Critical {C}urrent {D}ecay in {Josephson} {J}unctions {C}ontaining {A}ntiferromagnetic {NiMn}},\ }\href {https://doi.org/10.1109/TASC.2023.3257769} {\bibfield  {journal} {\bibinfo  {journal} {IEEE Trans. Appl. Supercond.}\ }\textbf {\bibinfo {volume} {33}},\ \bibinfo {pages} {1800903} (\bibinfo {year} {2023})}\BibitemShut {NoStop}%
\bibitem [{\citenamefont {Prateek}\ \emph {et~al.}(2024)\citenamefont {Prateek}, \citenamefont {Mechielsen}, \citenamefont {Hamida}, \citenamefont {Scholma}, \citenamefont {Junxiang},\ and\ \citenamefont {Aarts}}]{prateek2024fabrication}%
  \BibitemOpen
  \bibfield  {author} {\bibinfo {author} {\bibfnamefont {K.}~\bibnamefont {Prateek}}, \bibinfo {author} {\bibfnamefont {T.}~\bibnamefont {Mechielsen}}, \bibinfo {author} {\bibfnamefont {A.~B.}\ \bibnamefont {Hamida}}, \bibinfo {author} {\bibfnamefont {D.}~\bibnamefont {Scholma}}, \bibinfo {author} {\bibfnamefont {Y.}~\bibnamefont {Junxiang}},\ and\ \bibinfo {author} {\bibfnamefont {J.}~\bibnamefont {Aarts}},\ }\bibfield  {title} {\bibinfo {title} {Fabrication and properties of lateral {J}osephson junctions with a {RuO}$_2$ weak link},\ }\href@noop {} {\bibfield  {journal} {\bibinfo  {journal} {Supercond. Sci. Technol.}\ }\textbf {\bibinfo {volume} {37}},\ \bibinfo {pages} {035020} (\bibinfo {year} {2024})}\BibitemShut {NoStop}%
\bibitem [{\citenamefont {Esin}\ \emph {et~al.}(2026)\citenamefont {Esin}, \citenamefont {Kazmin}, \citenamefont {Barash}, \citenamefont {Timonina}, \citenamefont {Kolesnikov},\ and\ \citenamefont {Deviatov}}]{esin2026josephson}%
  \BibitemOpen
  \bibfield  {author} {\bibinfo {author} {\bibfnamefont {V.~D.}\ \bibnamefont {Esin}}, \bibinfo {author} {\bibfnamefont {D.~Y.}\ \bibnamefont {Kazmin}}, \bibinfo {author} {\bibfnamefont {Y.~S.}\ \bibnamefont {Barash}}, \bibinfo {author} {\bibfnamefont {A.~V.}\ \bibnamefont {Timonina}}, \bibinfo {author} {\bibfnamefont {N.~N.}\ \bibnamefont {Kolesnikov}},\ and\ \bibinfo {author} {\bibfnamefont {E.~V.}\ \bibnamefont {Deviatov}},\ }\bibfield  {title} {\bibinfo {title} {Josephson diode and spin-valve effects on the surface of altermagnet {CrS}b},\ }\href@noop {} {\bibfield  {journal} {\bibinfo  {journal} {JETP Lett.}\ }\textbf {\bibinfo {volume} {123}},\ \bibinfo {pages} {556} (\bibinfo {year} {2026})}\BibitemShut {NoStop}%
\bibitem [{\citenamefont {Jeon}\ \emph {et~al.}(2021)\citenamefont {Jeon}, \citenamefont {Hazra}, \citenamefont {Cho}, \citenamefont {Chakraborty}, \citenamefont {Jeon}, \citenamefont {Han}, \citenamefont {Meyerheim}, \citenamefont {Kontos},\ and\ \citenamefont {Parkin}}]{jeon2021long}%
  \BibitemOpen
  \bibfield  {author} {\bibinfo {author} {\bibfnamefont {K.-R.}\ \bibnamefont {Jeon}}, \bibinfo {author} {\bibfnamefont {B.~K.}\ \bibnamefont {Hazra}}, \bibinfo {author} {\bibfnamefont {K.}~\bibnamefont {Cho}}, \bibinfo {author} {\bibfnamefont {A.}~\bibnamefont {Chakraborty}}, \bibinfo {author} {\bibfnamefont {J.-C.}\ \bibnamefont {Jeon}}, \bibinfo {author} {\bibfnamefont {H.}~\bibnamefont {Han}}, \bibinfo {author} {\bibfnamefont {H.~L.}\ \bibnamefont {Meyerheim}}, \bibinfo {author} {\bibfnamefont {T.}~\bibnamefont {Kontos}},\ and\ \bibinfo {author} {\bibfnamefont {S.~S.~P.}\ \bibnamefont {Parkin}},\ }\bibfield  {title} {\bibinfo {title} {Long-range supercurrents through a chiral non-collinear antiferromagnet in lateral {J}osephson junctions},\ }\href@noop {} {\bibfield  {journal} {\bibinfo  {journal} {Nat. Mater.}\ }\textbf {\bibinfo {volume} {20}},\ \bibinfo {pages} {1358} (\bibinfo {year} {2021})}\BibitemShut {NoStop}%
\bibitem [{\citenamefont {Sürgers}\ \emph {et~al.}(2016)\citenamefont {Sürgers}, \citenamefont {Kittler}, \citenamefont {Wolf},\ and\ \citenamefont {Löhneysen}}]{surgers_2016}%
  \BibitemOpen
  \bibfield  {author} {\bibinfo {author} {\bibfnamefont {C.}~\bibnamefont {Sürgers}}, \bibinfo {author} {\bibfnamefont {W.}~\bibnamefont {Kittler}}, \bibinfo {author} {\bibfnamefont {T.}~\bibnamefont {Wolf}},\ and\ \bibinfo {author} {\bibfnamefont {H.~V.}\ \bibnamefont {Löhneysen}},\ }\bibfield  {title} {\bibinfo {title} {{Anomalous {H}all effect in the noncollinear antiferromagnet {M}n$_5${S}i$_3$}},\ }\href {https://doi.org/10.1063/1.4943759} {\bibfield  {journal} {\bibinfo  {journal} {AIP Adv.}\ }\textbf {\bibinfo {volume} {6}},\ \bibinfo {pages} {055604} (\bibinfo {year} {2016})}\BibitemShut {NoStop}%
\bibitem [{\citenamefont {Reichlova}\ \emph {et~al.}(2024)\citenamefont {Reichlova}, \citenamefont {Lopes~Seeger}, \citenamefont {Gonz{\'a}lez-Hern{\'a}ndez}, \citenamefont {Kounta}, \citenamefont {Schlitz}, \citenamefont {Kriegner}, \citenamefont {Ritzinger}, \citenamefont {Lammel}, \citenamefont {Leivisk{\"a}}, \citenamefont {Birk~Hellenes} \emph {et~al.}}]{reichlova_2024}%
  \BibitemOpen
  \bibfield  {author} {\bibinfo {author} {\bibfnamefont {H.}~\bibnamefont {Reichlova}}, \bibinfo {author} {\bibfnamefont {R.}~\bibnamefont {Lopes~Seeger}}, \bibinfo {author} {\bibfnamefont {R.}~\bibnamefont {Gonz{\'a}lez-Hern{\'a}ndez}}, \bibinfo {author} {\bibfnamefont {I.}~\bibnamefont {Kounta}}, \bibinfo {author} {\bibfnamefont {R.}~\bibnamefont {Schlitz}}, \bibinfo {author} {\bibfnamefont {D.}~\bibnamefont {Kriegner}}, \bibinfo {author} {\bibfnamefont {P.}~\bibnamefont {Ritzinger}}, \bibinfo {author} {\bibfnamefont {M.}~\bibnamefont {Lammel}}, \bibinfo {author} {\bibfnamefont {M.}~\bibnamefont {Leivisk{\"a}}}, \bibinfo {author} {\bibfnamefont {A.}~\bibnamefont {Birk~Hellenes}}, \emph {et~al.},\ }\bibfield  {title} {\bibinfo {title} {Observation of a spontaneous anomalous {H}all response in the {Mn}$_5${S}i$_3$ d-wave altermagnet candidate},\ }\href@noop {} {\bibfield  {journal} {\bibinfo  {journal} {Nat. Commun.}\ }\textbf {\bibinfo {volume} {15}},\ \bibinfo {pages} {4961} (\bibinfo {year}
  {2024})}\BibitemShut {NoStop}%
\bibitem [{\citenamefont {Kounta}\ \emph {et~al.}(2023)\citenamefont {Kounta}, \citenamefont {Reichlova}, \citenamefont {Kriegner}, \citenamefont {Lopes~Seeger}, \citenamefont {Bad'ura}, \citenamefont {Leiviska}, \citenamefont {Boussadi}, \citenamefont {Heresanu}, \citenamefont {Bertaina}, \citenamefont {Petit}, \citenamefont {Schmoranzerova}, \citenamefont {Smejkal}, \citenamefont {Sinova}, \citenamefont {Jungwirth}, \citenamefont {Baltz}, \citenamefont {Goennenwein},\ and\ \citenamefont {Michez}}]{kounta_2023}%
  \BibitemOpen
  \bibfield  {author} {\bibinfo {author} {\bibfnamefont {I.}~\bibnamefont {Kounta}}, \bibinfo {author} {\bibfnamefont {H.}~\bibnamefont {Reichlova}}, \bibinfo {author} {\bibfnamefont {D.}~\bibnamefont {Kriegner}}, \bibinfo {author} {\bibfnamefont {R.}~\bibnamefont {Lopes~Seeger}}, \bibinfo {author} {\bibfnamefont {A.}~\bibnamefont {Bad'ura}}, \bibinfo {author} {\bibfnamefont {M.}~\bibnamefont {Leiviska}}, \bibinfo {author} {\bibfnamefont {A.}~\bibnamefont {Boussadi}}, \bibinfo {author} {\bibfnamefont {V.}~\bibnamefont {Heresanu}}, \bibinfo {author} {\bibfnamefont {S.}~\bibnamefont {Bertaina}}, \bibinfo {author} {\bibfnamefont {M.}~\bibnamefont {Petit}}, \bibinfo {author} {\bibfnamefont {E.}~\bibnamefont {Schmoranzerova}}, \bibinfo {author} {\bibfnamefont {L.}~\bibnamefont {Smejkal}}, \bibinfo {author} {\bibfnamefont {J.}~\bibnamefont {Sinova}}, \bibinfo {author} {\bibfnamefont {T.}~\bibnamefont {Jungwirth}}, \bibinfo {author} {\bibfnamefont {V.}~\bibnamefont {Baltz}}, \bibinfo {author} {\bibfnamefont
  {S.~T.~B.}\ \bibnamefont {Goennenwein}},\ and\ \bibinfo {author} {\bibfnamefont {L.}~\bibnamefont {Michez}},\ }\bibfield  {title} {\bibinfo {title} {Competitive actions of {MnS}i in the epitaxial growth of {Mn}$_5${S}i$_3$ thin films on {S}i(111)},\ }\href {https://doi.org/10.1103/PhysRevMaterials.7.024416} {\bibfield  {journal} {\bibinfo  {journal} {Phys. Rev. Mater.}\ }\textbf {\bibinfo {volume} {7}},\ \bibinfo {pages} {024416} (\bibinfo {year} {2023})}\BibitemShut {NoStop}%
\bibitem [{SI()}]{SI}%
  \BibitemOpen
  \href@noop {} {}\bibinfo {note} {Supplemental Material available online at [URL]}\BibitemShut {NoStop}%
\bibitem [{\citenamefont {Wang}\ \emph {et~al.}(2012)\citenamefont {Wang}, \citenamefont {Pratt},\ and\ \citenamefont {Birge}}]{Wang_2012}%
  \BibitemOpen
  \bibfield  {author} {\bibinfo {author} {\bibfnamefont {Y.}~\bibnamefont {Wang}}, \bibinfo {author} {\bibfnamefont {W.~P.}\ \bibnamefont {Pratt}},\ and\ \bibinfo {author} {\bibfnamefont {N.~O.}\ \bibnamefont {Birge}},\ }\bibfield  {title} {\bibinfo {title} {Area-dependence of spin-triplet supercurrent in ferromagnetic {Josephson} junctions},\ }\href {https://doi.org/10.1103/PhysRevB.85.214522} {\bibfield  {journal} {\bibinfo  {journal} {Phys. Rev. B}\ }\textbf {\bibinfo {volume} {85}},\ \bibinfo {pages} {214522} (\bibinfo {year} {2012})}\BibitemShut {NoStop}%
\bibitem [{\citenamefont {Glick}\ \emph {et~al.}(2017)\citenamefont {Glick}, \citenamefont {Edwards}, \citenamefont {Korucu}, \citenamefont {Aguilar}, \citenamefont {Niedzielski}, \citenamefont {Loloee}, \citenamefont {Pratt}, \citenamefont {Birge}, \citenamefont {Kotula},\ and\ \citenamefont {Missert}}]{Glick2017triplet}%
  \BibitemOpen
  \bibfield  {author} {\bibinfo {author} {\bibfnamefont {J.~A.}\ \bibnamefont {Glick}}, \bibinfo {author} {\bibfnamefont {S.}~\bibnamefont {Edwards}}, \bibinfo {author} {\bibfnamefont {D.}~\bibnamefont {Korucu}}, \bibinfo {author} {\bibfnamefont {V.}~\bibnamefont {Aguilar}}, \bibinfo {author} {\bibfnamefont {B.~M.}\ \bibnamefont {Niedzielski}}, \bibinfo {author} {\bibfnamefont {R.}~\bibnamefont {Loloee}}, \bibinfo {author} {\bibfnamefont {W.~P.}\ \bibnamefont {Pratt}}, \bibinfo {author} {\bibfnamefont {N.~O.}\ \bibnamefont {Birge}}, \bibinfo {author} {\bibfnamefont {P.~G.}\ \bibnamefont {Kotula}},\ and\ \bibinfo {author} {\bibfnamefont {N.}~\bibnamefont {Missert}},\ }\bibfield  {title} {\bibinfo {title} {Spin-triplet supercurrent in {J}osephson junctions containing a synthetic antiferromagnet with perpendicular magnetic anisotropy},\ }\href {https://doi.org/10.1103/PhysRevB.96.224515} {\bibfield  {journal} {\bibinfo  {journal} {Phys. Rev. B}\ }\textbf {\bibinfo {volume} {96}},\ \bibinfo {pages} {224515}
  (\bibinfo {year} {2017})}\BibitemShut {NoStop}%
\bibitem [{\citenamefont {Barone}\ and\ \citenamefont {Patern\`{o}}(1982)}]{barone1982physics}%
  \BibitemOpen
  \bibfield  {author} {\bibinfo {author} {\bibfnamefont {A.}~\bibnamefont {Barone}}\ and\ \bibinfo {author} {\bibfnamefont {G.}~\bibnamefont {Patern\`{o}}},\ }\href {https://doi.org/10.1002/352760278X} {\emph {\bibinfo {title} {Physics and {A}pplications of the {J}osephson {E}ffect}}}\ (\bibinfo  {publisher} {John Wiley \& Sons, New York},\ \bibinfo {year} {1982})\BibitemShut {NoStop}%
\bibitem [{\citenamefont {Quarterman}\ \emph {et~al.}(2020)\citenamefont {Quarterman}, \citenamefont {Satchell}, \citenamefont {Kirby}, \citenamefont {Loloee}, \citenamefont {Burnell}, \citenamefont {Birge},\ and\ \citenamefont {Borchers}}]{quarterman2020distortions}%
  \BibitemOpen
  \bibfield  {author} {\bibinfo {author} {\bibfnamefont {P.}~\bibnamefont {Quarterman}}, \bibinfo {author} {\bibfnamefont {N.}~\bibnamefont {Satchell}}, \bibinfo {author} {\bibfnamefont {B.~J.}\ \bibnamefont {Kirby}}, \bibinfo {author} {\bibfnamefont {R.}~\bibnamefont {Loloee}}, \bibinfo {author} {\bibfnamefont {G.}~\bibnamefont {Burnell}}, \bibinfo {author} {\bibfnamefont {N.~O.}\ \bibnamefont {Birge}},\ and\ \bibinfo {author} {\bibfnamefont {J.~A.}\ \bibnamefont {Borchers}},\ }\bibfield  {title} {\bibinfo {title} {Distortions to the penetration depth and coherence length of superconductor/normal-metal superlattices},\ }\href {https://doi.org/10.1103/PhysRevMaterials.4.074801} {\bibfield  {journal} {\bibinfo  {journal} {Phys. Rev. Mater.}\ }\textbf {\bibinfo {volume} {4}},\ \bibinfo {pages} {074801} (\bibinfo {year} {2020})}\BibitemShut {NoStop}%
\bibitem [{\citenamefont {Sapkota}\ \emph {et~al.}(2026)\citenamefont {Sapkota}, \citenamefont {Sedai}, \citenamefont {Klaes}, \citenamefont {Loloee}, \citenamefont {Birge},\ and\ \citenamefont {Satchell}}]{sapkota_2026}%
  \BibitemOpen
  \bibfield  {author} {\bibinfo {author} {\bibfnamefont {A.}~\bibnamefont {Sapkota}}, \bibinfo {author} {\bibfnamefont {P.}~\bibnamefont {Sedai}}, \bibinfo {author} {\bibfnamefont {R.~M.}\ \bibnamefont {Klaes}}, \bibinfo {author} {\bibfnamefont {R.}~\bibnamefont {Loloee}}, \bibinfo {author} {\bibfnamefont {N.~O.}\ \bibnamefont {Birge}},\ and\ \bibinfo {author} {\bibfnamefont {N.}~\bibnamefont {Satchell}},\ }\bibfield  {title} {\bibinfo {title} {Large critical current density {J}osephson $\pi$-junctions with {PdN}i barriers},\ }\href@noop {} {\bibfield  {journal} {\bibinfo  {journal} {Appl. Phys. Lett.}\ }\textbf {\bibinfo {volume} {128}} (\bibinfo {year} {2026})}\BibitemShut {NoStop}%
\bibitem [{Dat()}]{Data}%
  \BibitemOpen
  \href@noop {} {}\bibinfo {note} {A. Sapkota \textit{et al.}, Rapid supercurrent decay in Mn$_5$Si$_3$ Josephson junctions, [Dataset], Texas Data Repository, 2026, https://doi.org/10.18738/T8/SRDQUU}\BibitemShut {NoStop}%
\end{thebibliography}%

\end{document}


\title{Supplemental Material for ``Rapid supercurrent decay in Mn$_5$Si$_3$ Josephson junctions''}

\author{Arjun Sapkota}
\affiliation{Materials Science, Engineering, and Commercialization Program, Texas State University, San Marcos, Texas 78666, USA}

\author{Kurt Lorenzen}
\affiliation{Department of Physics, Texas State University, San Marcos, Texas 78666, USA}

\author{Tyler Kuhn}
\affiliation{Department of Physics, Texas State University, San Marcos, Texas 78666, USA}

\author{Juan Gomez}
\affiliation{Department of Physics, Texas State University, San Marcos, Texas 78666, USA}

\author{Demet Korucu}
\affiliation{Department of Physics and Astronomy, Michigan State University, East Lansing, Michigan 48824, USA}

\author{Robert M. Klaes}
\affiliation{Department of Physics and Astronomy, Michigan State University, East Lansing, Michigan 48824, USA}

\author{Reza Loloee}
\affiliation{Department of Physics and Astronomy, Michigan State University, East Lansing, Michigan 48824, USA}

\author{Norman O. Birge}
\affiliation{Department of Physics and Astronomy, Michigan State University, East Lansing, Michigan 48824, USA}

\author{Nathan Satchell}
\email{satchell@txstate.edu}
\affiliation{Department of Physics, Texas State University, San Marcos, Texas 78666, USA}
\affiliation{Materials Science, Engineering, and Commercialization Program, Texas State University, San Marcos, Texas 78666, USA}

\vspace{24pt}

\maketitle

\section{Structural Characterization of a 40~nm film}

The \MnSi{} barriers in the Josephson junctions of the main text are too thin for structural characterization by x-ray diffraction. We therefore deposited a Pt(4)/\MnSi{}(40)/Pt(6) film, thicknesses in nm, under nominally identical conditions to the junction depositions, and characterized it by x-ray reflectivity (XRR) and x-ray diffraction (XRD) using a Rigaku SmartLab diffractometer with Cu K$\alpha$ radiation.

\begin{figure}[h]
    \centering
    \includegraphics[width=0.85\linewidth]{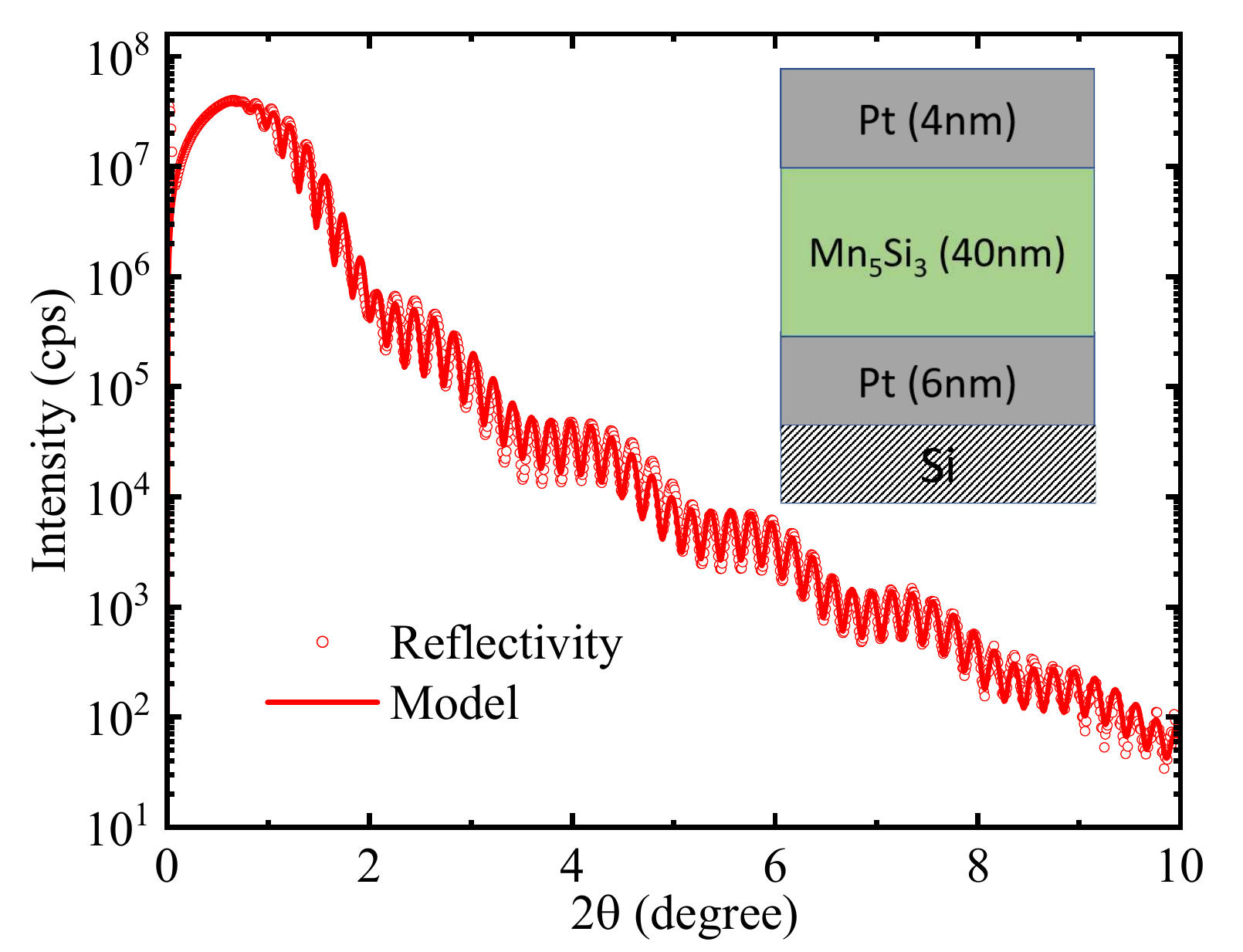}
    \caption{X-ray reflectivity of the Pt(4)/\MnSi{}(40)/Pt(6) film. Open symbols are the measured reflectivity and the solid line is the fit to the model given in Table~\ref{tab:S1}. Inset: the layer structure of the film, with nominal thicknesses.}
    \label{fig:S1}
\end{figure}

Figure~\ref{fig:S1} shows the XRR data together with a fit to a model of the layer structure. Fringes arising from the layer thicknesses are resolved to $2\theta = 10^{\circ}$. The parameters obtained from the fit are given in Table~\ref{tab:S1}.

\begin{table}[h]
\caption{\label{tab:S1}Parameters obtained from the fit to the x-ray reflectivity data of Figure~\ref{fig:S1}, alongside the nominal layer thicknesses.}
\begin{ruledtabular}
\begin{tabular}{lcccc}
Layer & Nominal (nm) & Thickness (nm) & Roughness (nm)  \\
\hline
Pt cap    & 4  & 2.90 & 0.22  \\
\MnSi{}   & 40 & 39.27 & 0.48  \\
Pt buffer & 6  & 5.30 & 0.33  \\
Substrate & -- & --   & 0.63  \\
\end{tabular}
\end{ruledtabular}
\end{table}

\begin{figure}[h]
    \centering
    \includegraphics[width=0.85\linewidth]{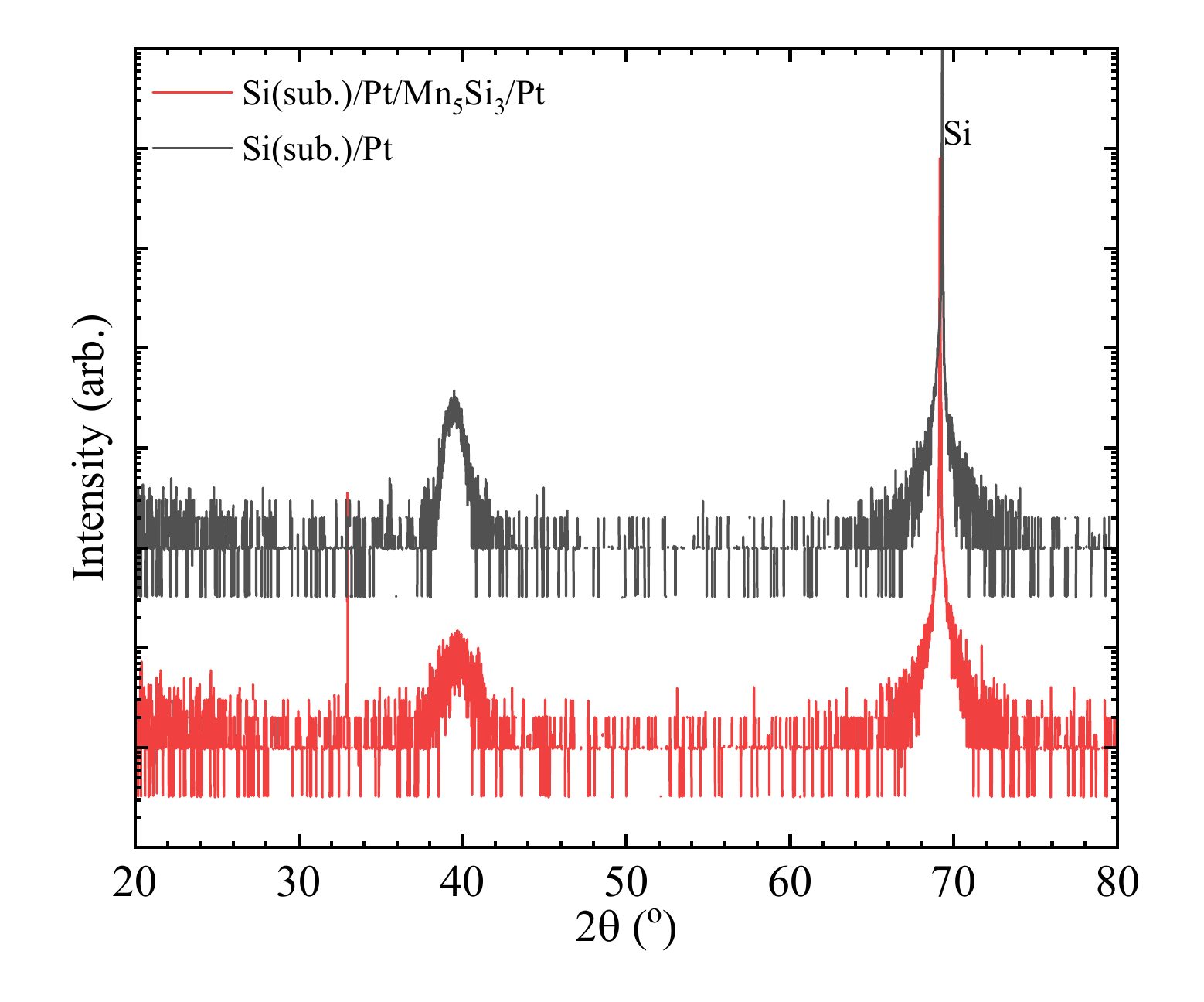}
    \caption{X-ray diffraction of the Pt(4)/\MnSi{}(40)/Pt(6) (red)  and Pt(8) (black) film on a Si/SiO$_2$ substrate. The film reflection at $2\theta = 39.6^{\circ}$ is consistent with both Pt(111) and \MnSi{}(210).}
    \label{fig:S2}
\end{figure}
Figure~\ref{fig:S2} shows a $\theta$--$2\theta$ XRD scan of the Pt reference sample and  Pt/\MnSi{}/Pt over the range $20^{\circ} \leq 2\theta \leq 80^{\circ}$. Aside from reflections arising from the Si substrate, both samples exhibit a diffraction peak at at $2\theta = 39.6^{\circ}$, consistent with both Pt(111) and \MnSi{}(210). The observation of this peak in the single layer Pt reference sample indicates that it is most likely Pt(111).

\bibliography{Refs}